\documentclass[runningheads]{llncs}
\usepackage{bm}
\usepackage{amsmath,amssymb}
\usepackage{graphicx}
\usepackage{graphicx}
\usepackage{tikz}
\usepackage{pgfplots}
\pgfplotsset{compat=1.18}

\usepackage[utf8]{inputenc}
\usepackage{newunicodechar}
\usepackage{booktabs}
\newunicodechar{【}{[}
\newunicodechar{】}{]}

\title{Intertemporal Pricing of Time-Bound Stablecoins:\\ Measuring and Controlling the Liquidity-of-Time Premium}
\titlerunning{Intertemporal Pricing of Time-Bound Stablecoins}
\author{Ailiya Borjigin \and Cong He}
\institute{Probe Group Pte. Ltd., Singapore\\
\email{\{Ailiya,~CongHe\}@probe-group.com}}
\begin{document}
\maketitle

\begin{abstract}
Time-bound stablecoins are a new class of decentralized finance (DeFi) assets that temporarily tokenize traditional securities during market off-hours, enabling continuous cross-market liquidity. This paper introduces the \emph{Liquidity-of-Time Premium (TLP)} as the extra return or cost associated with providing liquidity when an asset’s primary market is closed. We develop a theoretical pricing model for time-bound stablecoins that defines a no-arbitrage band for their values across different expiry horizons, and we design a dynamic risk control mechanism that adjusts loan-to-value (LTV) ratios in real time to keep the TLP within a target band. Our analysis combines financial engineering (no-arbitrage conditions, option-theoretic pricing) with empirical finance (event studies on cross-listed stocks and futures) to measure TLP under real-world time-zone frictions. We formally define TLP and derive closed-form expressions for its term structure under idealized conditions, and we simulate various scenarios to illustrate how TLP behaves under different volatility and collateralization settings. We further propose a dynamic LTV adjustment policy whereby the protocol raises or lowers collateral requirements to supply or curtail time-bound stablecoin liquidity, analogous to a central bank adjusting interest rates to stabilize a currency peg. Empirical proxies for TLP are discussed – including American Depositary Receipt (ADR) premiums, overseas index futures vs.\ cash index divergence, and pre-market vs.\ official close price gaps – to demonstrate how TLP can be identified and quantified from market data. The results show that TLP tends to increase with the length of market closure and asset volatility, but can be effectively constrained by adaptive LTV policies. We present simulation backtests and illustrative figures – such as the TLP term structure curve, capital efficiency vs.\ tail-risk trade-off, and a time--liquidity heatmap – to visualize key findings. This work positions time-bound stablecoins as a mechanism to arbitrage away temporal market inefficiencies, and offers guidance on protocol design (vault structure, oracle anchoring to closing prices, on-chain auction liquidation) to safely realize a global 24/7 trading loop. Finally, we survey related literature in finance (market microstructure of overnight trading, cross-market arbitrage) and DeFi (stablecoin design, liquidation auctions), and conclude with implications for market efficiency and future research directions.
\end{abstract}

\section{Introduction}
Global financial markets have long been segmented by trading hours and time zones, leaving substantial value dormant when local exchanges close. For example, a stock listed in New York cannot be traded or reallocated when the U.S. market is closed, even if markets in Asia or Europe are open. This mismatch creates “liquidity gaps” in time: investors are unable to respond to news or arbitrage price discrepancies across regions until each market’s next opening. Research has documented significant price movements and risk premia associated with these overnight periods. In U.S.\ equities, a large portion of the equity risk premium is earned overnight when other markets (e.g., Europe) are active【8】. Cross-listed stocks often exhibit asynchronous price discovery, with assets like American Depositary Receipts (ADRs) in New York impounding news while the home market is closed【7】. These effects underscore that time---specifically, the lack of simultaneous trading---commands a premium or cost that market participants implicitly pay【1,2,3,4,5】.

Recent developments in both traditional finance and DeFi point toward a convergence on 24/7 market access. Major exchanges have begun extending trading hours or launching around-the-clock sessions, and brokers now offer 24-hour trading in hundreds of equities to meet investor demand. In parallel, DeFi has enabled continuous trading of crypto assets, inspiring efforts to bring real-world assets on-chain for always-on liquidity. However, existing approaches face limitations. After-hours markets are often illiquid and fragmented, and traditional solutions like margin loans or repurchase agreements (repos) to bridge overnight gaps are accessible only to select players and can be costly. Early attempts to tokenize stocks in DeFi have required centralized price oracles or lacked robust risk controls, undermining their trustworthiness.

The \emph{Smart-Stable Securities (SSS) Protocol} is a recently proposed DeFi mechanism that directly tackles the time-gap problem by introducing time-bound stablecoins fully backed by off-hour securities. Under SSS, an investor can lock a security (such as a stock) after market close and mint a stablecoin equal to the stock’s value at the last official closing price. This stablecoin remains fixed in value (pegged 1:1 to the security’s closing price) during the closed period and expires at the next market open, at which point the investor must redeem it by returning the same shares originally locked. Effectively, SSS creates a short-term stock-backed loan that converts idle shares into deployable capital overnight. When the market reopens, the loan is reversed: the investor returns the stablecoin and retrieves the share (or, if they fail to repay, the share is liquidated to cover the stablecoin). This design allows an investor to immediately reinvest the value of a closed-market asset into another open market, achieving a continuous 24-hour capital rotation.

While the SSS protocol provides the infrastructure for bridging liquidity across time zones, a core question remains: \emph{How should these time-bound stablecoins be priced, and what premium reflects the value of liquidity over time?} We define this Liquidity-of-Time Premium (TLP) as the difference between the stablecoin’s fair value during the closed period and the underlying asset’s stale price, representing the market’s valuation of immediacy versus waiting. Intuitively, TLP captures how much extra “interest” a borrower is willing to pay (or a lender demands) to unlock liquidity now rather than later. Our work provides a formal treatment of TLP, establishing a no-arbitrage band for stablecoin prices and showing how arbitrage opportunities arise if TLP strays outside that band.

Crucially, because no immediate cross-market arbitrage is possible when one market is closed, the usual law of one price is relaxed within limits. Price discrepancies can emerge---for instance, an ADR might trade at a premium to the last local closing price if positive news breaks overnight---but these discrepancies are bounded by the risks and costs of arbitrageurs who plan to profit when markets realign. We derive conditions for these bounds, showing that time-bound stablecoins should trade within a “no-arbitrage corridor” determined by factors such as expected price movement, volatility, and transaction costs. If the stablecoin’s price is too low relative to the underlying (reflecting a high TLP), arbitrageurs could buy the stablecoin cheaply and anticipate redeeming it for full value at open, essentially earning an excess return. Conversely, if the stablecoin’s price is too high (negative TLP), it would be attractive for investors to mint and sell stablecoins (by locking collateral) to capture that premium, increasing supply until the price normalizes.

The second major contribution of this paper is a dynamic risk management framework for \emph{controlling} TLP via the protocol’s parameters. The SSS design already incorporates a dynamic LTV system---essentially, an adaptive collateralization ratio---based on the asset’s risk profile. We extend this idea by treating LTV as a policy lever to target a desired TLP range. If TLP widens (the stablecoin trades at a significant discount, indicating high demand for liquidity or high perceived risk), the protocol can respond by reducing the LTV (requiring more collateral per stablecoin). A lower LTV curtails the supply of stablecoins and increases their backing, which should compress the TLP by making the stablecoin safer and scarcer. Conversely, if the stablecoin consistently trades at or above par (TLP near zero or negative), the protocol could cautiously raise LTV to increase capital efficiency for users, expanding supply until a modest positive TLP reappears as compensation for overnight risk. This feedback mechanism resembles a monetary authority adjusting interest rates to maintain a currency peg: here the “peg” is the 1:1 parity between the stablecoin’s face value and the underlying’s closing price, and TLP plays the role of a short-term interest rate differential.

In the remainder of the paper, we develop these points rigorously. \textbf{Section 2 (\textit{Related Work}) }reviews literature on cross-market price discovery, overnight risk premia, and existing stablecoin mechanisms. \textbf{Section 3 (\textit{Protocol Overview}) }summarizes the SSS protocol’s architecture and terminology---including time-bound stablecoins, vaults, oracle pricing at official closes, equivalent-share return, and on-chain liquidations---to ensure the model is well-grounded in the actual mechanism design. \textbf{Section 4 (\textit{Theoretical Model})} formally defines the Liquidity-of-Time Premium and derives a no-arbitrage pricing band for time-bound stablecoins across different time-to-expiry durations. We model the stablecoin’s payoff as a contingent claim on the underlying asset’s opening price, showing that the stablecoin’s valuation can be viewed as the underlying’s closing price minus an embedded put option (reflecting the borrower’s option to default and surrender collateral). This formulation yields closed-form expressions for TLP in terms of option Greeks under lognormal assumptions for price gaps, and it illuminates how TLP grows with volatility and gap length. We then describe a control policy wherein LTV is adjusted dynamically in response to observed TLP deviations, and we prove a proposition that this policy can keep TLP within a target interval under certain conditions. \textbf{Section 5 (\textit{Empirical Identification})} proposes strategies for measuring TLP using market data. Direct observations of time-bound stablecoins are not yet available (as SSS is an emerging concept), so we turn to analogs: ADR price deviations from local shares, overnight index futures vs.\ next-day index opens, and pre-market trading vs.\ prior close in equities. For instance, the ADR of a stock often trades at a premium or discount relative to the last local close---that spread, adjusted for any FX changes, is essentially a realized Liquidity-of-Time Premium that arbitrageurs aim to capture. We outline an event-study methodology where major after-hours news (earnings surprises, geopolitical events) are used to induce large TLPs, and we measure how quickly these premiums dissipate once both markets open. Additionally, we discuss the use of opening auction data (which often include price gaps) to infer the cost of transacting at the open versus prior close, providing another window into TLP. \textbf{Section 6 (\textit{Simulation \& Backtest}) }presents simulations of an SSS-like system to illustrate the dynamic between LTV, volatility, and TLP. We build an agent-based model where borrowers and arbitrageurs interact: borrowers decide how much to mint based on cost (TLP and fees), and arbitrageurs trade the stablecoin and underlying asset based on arbitrage incentives. We simulate various scenarios (including stress events) and show how a dynamic LTV policy can stabilize TLP. We also perform a historical backtest using past market data to estimate how the system might have behaved if it were live in recent years. \textbf{Section 7 (\textit{Design Implications})} discusses practical considerations for implementing time-bound stablecoins, such as vault management, oracle design (anchoring to official close prices), and liquidation mechanism choices (on-chain auction vs.\ off-chain execution). We draw parallels to existing stablecoin protocols and risk management practices. Finally, \textbf{Section 8 (\textit{Conclusion}) }concludes with implications for market efficiency and future research directions.

\section{Related Work}
\subsection{Cross-Market Price Discovery}
Our study of liquidity across time builds on a rich literature examining how information is impounded into prices across markets that do not trade simultaneously. Classic works by Eun and Shim【1】 and Hamao et al.【2】 first documented the international transmission of stock market movements and volatility through non-concurrent trading sessions. Subsequent research has explored both overlapping and sequential trading regimes. For example, Gagnon and Karolyi【3】 analyze cross-market arbitrage when markets for a stock trade in multiple time zones, and Grammig et al.【4】 examine price discovery for internationally cross-listed stocks during overlapping trading hours (accounting for exchange rate effects). Cross-listed stocks with non-overlapping trading hours provide natural experiments in asynchronous trading. Alsayed and McGroarty【5】 showed that arbitrage between stocks and their New York ADRs yields only small per-trade profits but accumulates to substantial returns, indicating a strong tendency toward price parity due to two-way convertibility. Bogomolov et al.【6】 focused on Asia-Pacific stocks with ADRs and found economically significant overnight arbitrage opportunities, even when execution is delayed until the home market reopens. Their work formalized the idea of a mean-reverting price spread between markets, and demonstrated that while deviations from parity can be exploited, the trades involve bearing overnight risk. We generalize this concept by defining TLP as the required premium for bearing exactly that overnight risk. In effect, the arbitrage profits documented in these studies are the reward for providing liquidity when one side of the market is closed---in other words, an empirical measure of TLP.

Another relevant aspect is the circadian mismatch in trader behavior. Dickinson et al.【10】 find that traders operating outside their usual hours (for instance, U.S.\ traders active during Asian market hours) may be less efficient or face higher cognitive costs, leading to mispricings. This suggests that even if markets were open 24/7, human and institutional constraints could result in thinner liquidity or less efficient prices during “unnatural” trading hours. Our TLP concept can thus be interpreted as compensating not only for time-zone constraints but also for the lower liquidity and higher uncertainty in off-hours periods. 

\subsection{Overnight Returns and Liquidity Frictions}
Equity markets exhibit the so-called “overnight drift” phenomenon---stock returns between the previous close and the next opening are on average positive, even in the absence of broad news【8】. There is evidence that these overnight returns often partially reverse during the next trading day, especially following large overnight moves【8】. Some studies attribute the overnight return premium to liquidity imbalances and risk aversion on the part of market makers at the close. For example, Nagel【9】 proposes that dealers demand a premium (or mark down prices before close) to compensate for the risk of holding inventory overnight when liquidity is lower. By the open, as liquidity resurfaces, prices rebound and the previous imbalance is corrected, yielding a reversal. These findings buttress our focus on the liquidity-of-time: they indicate a structural return component tied to the close-to-open interval, reflecting compensation for liquidity provision over time.

From a market microstructure perspective, after-hours trading is generally associated with wider bid-ask spreads and thinner order books. Madhavan【11】 models how liquidity supply is lower in off-hours, leading to larger price impacts per trade; this translates to a cost of immediacy that is closely related to TLP. Empirically, Barclay and Hendershott【12】 found that price discovery still occurs after hours but at a slower rate and with higher transaction costs, as only certain informed or liquidity-motivated traders participate. These insights imply that a mechanism which provides additional liquidity during normally closed periods (like time-bound stablecoins) could improve overall market efficiency by narrowing spreads and arbitrage gaps that would otherwise persist until the open.

\subsection{Stablecoins and Decentralized Liquidity}
Turning to the DeFi arena, our work intersects with research on stablecoins and on-chain risk management. A growing body of literature examines how stablecoin pegs are maintained via collateralization and algorithmic control. Kozhan and Viswanath-Natraj【13】 analyze crypto-backed over-collateralized stablecoins (e.g., DAI), highlighting how collateral volatility and liquidation mechanisms drive stablecoin stability and collateral risk. Mita et al.【30】 provide a taxonomy of stablecoins based on their collateral nature (fiat-backed, crypto-backed, algorithmic) and analyze their price stabilization mechanisms through economic lenses. These works underscore that over-collateralization and timely liquidations can effectively preserve a peg, albeit at the cost of capital efficiency. Qin et al.【14】 empirically study DeFi liquidation events and find that fixed-margin liquidation designs (like those in MakerDAO) can favor liquidators over borrowers, revealing instabilities when collateral prices gap down. Their findings point to the importance of adaptive risk controls in lending protocols. 

Our paper synthesizes these insights at a new intersection: a stablecoin that exists purely to arbitrage time, and a framework to price and stabilize it. We draw on concepts from both TradFi and DeFi. For instance, the dynamic LTV adjustment in our model is akin to margin requirement adjustments by brokers or clearinghouses to account for “gap risk”【11】. Likewise, the on-chain liquidation approach in SSS is inspired by proven designs in protocols like MakerDAO【16】, which rely on over-collateralization and automated auctions to maintain solvency. By explicitly pricing the liquidity-of-time and actively managing it via protocol parameters, our work extends stablecoin design into the temporal domain. In essence, we treat time as the underlying asset whose liquidity is being tokenized, and integrate traditional financial engineering (no-arbitrage pricing) with decentralized mechanisms (smart contracts, oracles, on-chain auctions) to ensure that this new form of stablecoin remains stable and useful.

\section{Protocol Overview}
The SSS protocol provides the infrastructure enabling time-bound stablecoins. We briefly outline its key components and terminology (adopting descriptions from the SSS whitepaper【25】):

\paragraph{Time-Bound Stablecoin.} A short-term stablecoin token minted against a specific security after its market close and valid only until the next market opening. It is pegged 1:1 to the security’s value at the last close (in the local currency of that market) and is designed not to fluctuate during the overnight period. For example, locking \$10,000 worth of a stock at NYSE close yields 10,000 units of a USD-denominated stablecoin (e.g., SSS-USD). This token remains at \$1.00 value throughout the night by construction – the protocol’s price oracle will not update the price until NYSE reopens. By automatically expiring and being redeemed at the next open, the stablecoin avoids long-term divergence from the underlying; it functions essentially as an overnight collateralized loan. The moment the stock’s market opens, the stablecoin is scheduled for redemption and destruction.

\paragraph{Vaults and Equivalent-Share Return.} When a user mints time-bound stablecoins, they deposit their shares into an on-chain vault smart contract. The vault holds the collateral securely (likely via a linkage to a qualified custodian who actually holds the real shares) and issues the stablecoins. A fundamental rule is the \emph{equivalent-share return} requirement: the same number of shares that were locked must be returned by the borrower upon redemption. In other words, you get your exact shares back only if you repay all the stablecoins; if you default, those shares are liquidated to make stablecoin holders whole. This ensures continuity of ownership and prevents any attempt to substitute different collateral or avoid returning shares. Essentially, each stablecoin is fully backed by specific shares; if the borrower fails to return the stablecoins, the protocol will auction off those shares to redeem the stablecoins. The vault also tracks any accrued fees or interest that the borrower owes for the duration of the loan.

\begin{figure}[t]
  \centering

  \begin{tikzpicture}[x=0.35cm,y=0.5cm]
    \draw[thick] (0,0) -- (24,0);
    \foreach \h in {0,4,8,12,16,20,24} {
      \draw (\h,0) -- (\h,-.3) node[below]{\small \h};
    }
    \node[below right] at (0,-1.3) {\small Hours (UTC)};

    \node[left] at (-.5,3) {\small Asia};
    \fill[gray!20] (0,2.5) rectangle (24,3.5);
    \fill[gray!80] (0,2.5) rectangle (8,3.5);     
    \fill[gray!80] (12,2.5) rectangle (16,3.5);   

    \node[left] at (-.5,2) {\small Europe};
    \fill[gray!20] (0,1.5) rectangle (24,2.5);
    \fill[gray!80] (8,1.5) rectangle (16,2.5);

    \node[left] at (-.5,1) {\small US};
    \fill[gray!20] (0,.5) rectangle (24,1.5);
    \fill[gray!80] (14,.5) rectangle (21,1.5);

    \draw[->,thick,blue] (7,3) -- (9,2);  
    \draw[->,thick,blue] (15,2) -- (16,1); 
    \draw[->,thick,blue] (21,1) -- (23,3); 

  \end{tikzpicture}

  \caption{Conceptual time–liquidity heatmap across regions. Dark bands indicate open sessions; arrows indicate SSS bridging during closures. }
  \label{fig:time_liquidity_heatmap}
\end{figure}
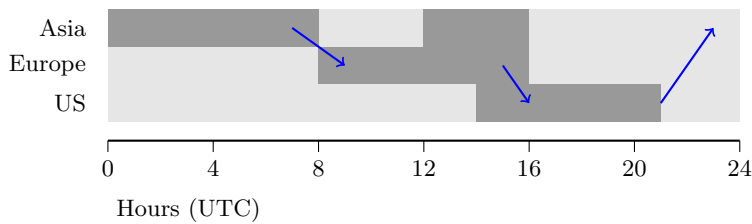

\paragraph{Oracle Anchoring to Official Closing Prices.} SSS uses a decentralized price oracle that reports the official closing price of the underlying asset (and later the official opening price). After the market closes, the oracle publishes the closing price which becomes the peg for the stablecoin’s value. Importantly, during the overnight period, the oracle does not update the price even if related markets (like futures or overseas listings) move – the stablecoin’s on-chain value remains anchored to the last official close. When the market reopens, the oracle updates to the opening price. This design choice means the stablecoin is effectively “stale-priced” overnight, simplifying valuation: 1 stablecoin token represents 1 share’s worth of value at last close. It also means that any divergence between this peg and an updated fundamental value (due to news) is reflected as a premium/discount in the stablecoin’s secondary market price (which is exactly TLP).

\paragraph{LTV and Fees.} Similar to a lending protocol, SSS defines a Loan-to-Value ratio at issuance (LTV$_c$) that governs how many stablecoins can be minted per value of collateral. For example, an LTV of 0.9 means a user locking \$100 of stock can mint at most 90 stablecoin tokens (each worth \$1 at close). In practice, SSS may allow up to LTV = 1.0 (100\% of collateral value) for high-quality assets, since the loan is very short-term. However, to account for overnight risk, a slightly lower LTV or a safety buffer might be used. The protocol can charge a small fee or interest for the service (e.g., 0.1\% of the stablecoin amount), accruing to a treasury or insurance fund. This fee compensates liquidity providers and can be adjusted based on demand (much like interest rates in money markets).

\paragraph{Liquidation Mechanism.} If a borrower fails to redeem by the deadline (just before market open) or if a severe price drop in the collateral occurs that makes the position unsafe, a liquidation process is triggered. Since the underlying market is closed during the stablecoin’s life, immediate on-chain liquidation is challenging. The simpler design (as implied by the SSS whitepaper) is that liquidation occurs at or immediately after the market reopens. For instance, if the stock drops significantly overnight (e.g., due to news pre-market) such that the collateral value falls below the stablecoin amount, the protocol will, at open, auction or sell the shares to cover the stablecoin liability. An on-chain auction can be implemented via smart contract: participants bid stablecoins (or other accepted currency) for the shares. A Dutch auction format might be used, starting at the last close price and descending until buyers step in. This ensures that stablecoin holders can be made whole (by buying back and burning the undercollateralized stablecoins using the auction proceeds). Any shortfall would be absorbed by an insurance fund or the protocol, while any surplus (if collateral sells for more than needed) could go to the borrower or insurance fund.

Overall, the SSS protocol provides a novel asset transformation: essentially a swap with one static leg and one active leg. One leg is a locked asset whose price is static (the closed-market stock), and the other leg is a liquid token that can move (the stablecoin). The design is fully collateralized and self-contained, with formal safeguards like equivalent-share return and oracle anchoring to maintain trust. In the next section, we assume this framework (one unit of asset locked, stablecoins minted, etc.) and analyze pricing under these rules. The mechanics described above ensure that at any time during the stablecoin’s life, its redemption value is well-defined and fully backed, which is a prerequisite for rigorous pricing analysis.

\section{Theoretical Model}
\subsection{Liquidity-of-Time Premium Definition}

We begin by formally defining the Liquidity-of-Time Premium in the context of time-bound stablecoins. Consider an asset (stock) with price $S_c$ at the close of the primary market ($c$ for close). An investor locks one share of this stock and mints one stablecoin with face value $S_c$. At the next opening ($o$ for open), the stock’s price is $S_o$. The stablecoin holder is entitled to redeem for the share (worth $S_o$) if the borrower defaults, or receive $S_c$ in cash if the borrower repays. Essentially, the stablecoin is a claim that will be worth $\min(S_c,\;S_o)$ at $o$ (assuming the protocol triggers default if $S_o < S_c$, since otherwise the borrower would repay to reclaim the share).

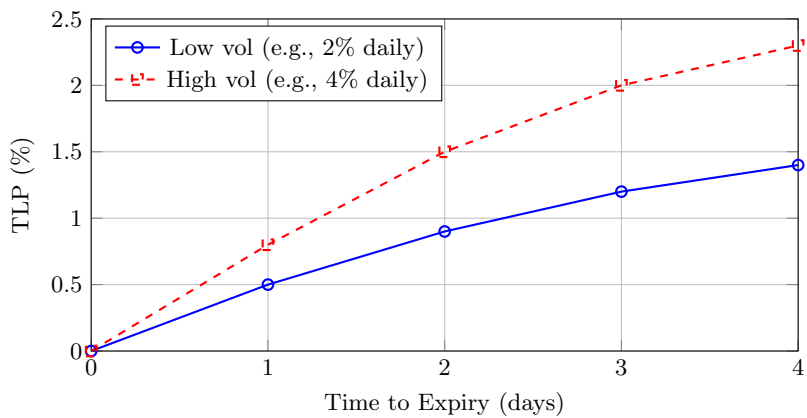
\begin{figure}[t]
  \centering

  \begin{tikzpicture}
    \begin{axis}[
      width=.9\linewidth,height=6cm,
      xlabel={Time to Expiry (days)},
      ylabel={TLP (\%)},
      xmin=0,xmax=4,
      ymin=0,ymax=2.5,
      xtick={0,1,2,3,4},
      ytick={0,0.5,1,1.5,2,2.5},
      legend style={at={(0.02,0.98)},anchor=north west,font=\small},
      grid=both
    ]
      \addplot+[mark=o,thick] coordinates {
        (0,0) (1,0.5) (2,0.9) (3,1.2) (4,1.4)
      }; \addlegendentry{Low vol (e.g., 2\% daily)}

      \addplot+[mark=square,thick,dashed] coordinates {
        (0,0) (1,0.8) (2,1.5) (3,2.0) (4,2.3)
      }; \addlegendentry{High vol (e.g., 4\% daily)}
    \end{axis}
  \end{tikzpicture}

  \caption{Term structure of the Liquidity-of-Time Premium (TLP) across 1–3 day closures for assets with different daily volatility. }
  \label{fig:tlp_term_structure}
\end{figure}

We can express this payoff as:
\[ \min(S_c,\;S_o) \;=\; S_c \;-\; \max(0,\;S_c - S_o)~, \] 
which shows that the stablecoin’s payoff equals the full face value $S_c$ minus the payoff of a put option (with strike $S_c$) on the stock’s $o$-price. In other words:
\[ \text{Stablecoin value at issuance} \;=\; S_c \;-\; P(S_c), \] 
where $P(S_c)$ is the time-$c$ value of a put option expiring at $o$ with strike $S_c$. This simple but powerful result indicates that the stablecoin is essentially equivalent to a fully collateralized claim \emph{minus} a default option (the put option represents the borrower’s option to surrender the collateral if its value falls). The Liquidity-of-Time Premium (TLP) can be defined as the discount of the stablecoin’s fair price $P_c$ below $S_c$:
\[ \text{TLP} = \frac{S_c - P_c}{S_c}~, \] 
which, in this idealized setting, would equal $\frac{P(S_c)}{S_c}$ (the percentage value of that put option). Intuitively, TLP is zero if there is no anticipated overnight risk or illiquidity (stablecoin trades at par $S_c$), and positive if the stablecoin is expected to pay out less than $S_c$ in expectation due to default risk or illiquidity (hence it trades at a discount). In the absence of any frictions or risks, $P(S_c)=0$ and TLP would be 0 (no premium for liquidity). As volatility or default risk increases, the put option value grows, and so does TLP.

Under a simple model where the overnight return $R = S_o/S_c - 1$ is lognormally distributed with volatility $\sigma$ and zero drift (risk-neutral), one can derive a closed-form for the put price using the Black-Scholes formula (treating the overnight as a short maturity). For instance, if $\tau$ is the overnight duration in years (e.g., $\tau\approx 1/252$ for one trading day) and assuming no interest/dividends for simplicity, we have:
\[ P(S_c) = S_c \Phi(-d_2) - S_c e^{-\sigma^2 \tau/2} \Phi(-d_1)~, \] 
where $d_{1,2} = \frac{\ln(1) \pm (\sigma^2\tau/2)}{\sigma\sqrt{\tau}} = \pm \frac{\sigma\sqrt{\tau}}{2}$, and $\Phi$ is the standard normal CDF. This yields a positive $P(S_c)$ increasing in $\sigma\sqrt{\tau}$. For small $\tau$, $P(S_c) \approx \frac{1}{2}S_c (\sigma\sqrt{\tau})$ (scaling with volatility), implying TLP $\approx \frac{1}{2}\sigma\sqrt{\tau}$ in this rough approximation. Thus, TLP increases with the length of the closed period and the asset’s volatility.

\subsection{No-Arbitrage Pricing Band}
Because cross-market arbitrage cannot be executed until the market reopens, the stablecoin’s price $P_c$ at issuance can deviate from $S_c$ within certain bounds. We consider a risk-neutral arbitrageur with access to both the stock and stablecoin markets and derive a no-arbitrage band.

If the stablecoin is \emph{undervalued} (trading at a discount too deep), say $P_c < S_c - \Delta$, an arbitrage strategy would be: buy stablecoins at $P_c$ and simultaneously \emph{short-sell} the underlying stock (or equivalently, defer buying the stock until open). At $o$, use the stock (obtained from covering the short or buying at open) to redeem stablecoins for $S_c$ worth of stock or cash. The payoff would be $S_c - P_c -$ (short financing cost). The risk is that $S_o \neq S_c$; the arbitrageur’s stock short is covered at $S_o$, and the stablecoin redemption yields either stock (if borrower defaults) or cash $S_c$ (if borrower repays). In either case, the arbitrageur ends up with either $S_c - P_c$ (if borrower repays in cash $S_c$ and stock short is closed at $S_o = S_c$) or $S_o - P_c - (S_o - S_c)$ (if borrower defaults and arbitrageur gets stock worth $S_o$ but owes $S_o$ on short; after closing, net $S_c - P_c$). Thus ignoring financing and assuming the arbitrageur can stomach the overnight position, the expected profit (under risk-neutral pricing) would be $S_c - P_c$. For no-arbitrage in expectation, we require $P_c \ge \mathbb{E}[ \min(S_c, S_o) ]$. More directly, considering the default probability $\pi = \Pr(S_o < S_c)$ and expected loss fraction $\ell = \mathbb{E}[ (S_c - S_o)/S_c \mid S_o < S_c ]$, the fair pricing would satisfy:
\[ P_c = (1-\pi) S_c + \pi (S_c - \ell S_c) = S_c - \pi \ell S_c. \]
This yields a lower bound: $P_c \ge S_c - \pi \ell S_c$, otherwise an arbitrageur would expect positive profit. Conversely, if the stablecoin is \emph{overvalued} ($P_c > S_c$), one could arbitrage by borrowing stablecoins (minting by locking shares) and selling them for $P_c$, then unwinding at open by buying back with the share. However, because one cannot mint more than $S_c$ per share and must lock collateral, $P_c$ cannot rationally exceed $S_c$ (that would be an immediate arbitrage: lock \$100 of stock, get 100 stablecoins, sell them for >\$100, and later repay with the share). Thus the upper bound is $P_c \le S_c$. In summary, the no-arbitrage band is:
\[ S_c - \pi \ell S_c \;\le\; P_c \;\le\; S_c~, \] 
where $\pi \ell S_c$ represents the risk discount for illiquidity and default. As market frictions vanish (e.g., if one could continuously trade or perfectly hedge overnight moves), we have $\pi \to 0$ or $\ell \to 0$ and the band collapses with $P_c \to S_c$. Conversely, higher volatility or longer closures increase $\pi$ and $\ell$, widening the band and allowing a larger TLP.

\subsection{Dynamic TLP Control via LTV Adjustments}
A unique aspect of our approach is using protocol parameters to actively manage TLP. The primary lever is the Loan-to-Value ratio (LTV) allowed for minting. A higher LTV means borrowers can mint more stablecoins per collateral value, increasing stablecoin supply relative to locked collateral; a lower LTV restricts supply, making stablecoins scarcer. The protocol’s objective is to keep the stablecoin price at parity ($P_c \approx S_c$, i.e., TLP $\approx 0$) within a narrow band, to ensure confidence in the stablecoin.

We can frame this as an optimal control problem: maximize liquidity provision (LTV high) while maintaining default risk below a threshold. For a given asset volatility and closure duration, default probability is $\pi = \Pr(S_o < \text{LTV} \cdot S_c)$. Assuming lognormal returns, $\pi = \Phi\!\Big(\frac{\ln(\text{LTV}) + \mu\tau}{\sigma\sqrt{\tau}}\Big)$ under risk-neutral drift $\mu$. For a target maximal default probability $\epsilon$ (e.g., 1\%), one could solve for the maximum LTV that satisfies $\pi \le \epsilon$. This would be $\text{LTV}^{*} = e^{-\mu\tau + \sigma \sqrt{\tau} \Phi^{-1}(\epsilon)}$. If that LTV is too restrictive (very low), the protocol might accept a slightly higher $\epsilon$ or maintain an insurance fund to absorb rare losses.

The key idea of dynamic control is to adjust LTV in real time based on observed stablecoin price (TLP). If the stablecoin trades below peg (indicating TLP is positive and possibly widening), the protocol would reduce LTV for new issuances and perhaps even require existing borrowers to partially repay (or add collateral) if feasible. This contraction of supply should raise the stablecoin price (reduce TLP) by making liquidity more scarce. If instead the stablecoin persistently trades at or above parity (TLP $\le 0$), the protocol can afford to raise LTV gradually, allowing more liquidity until a slight discount appears. This is analogous to a central bank adjusting interest rates to keep a currency peg within a band.

Formally, suppose the protocol observes a stablecoin market price $P_c^{\text{obs}}$. Define TLP$_{\text{obs}} = (S_c - P_c^{\text{obs}})/S_c$. We choose an upper and lower bound $\{\text{TLP}_{\min},\,\text{TLP}_{\max}\}$ as the acceptable range (e.g., 0 to 1\% discount). We then set a rule: 
\[ \Delta \text{LTV} = -k \,(\text{TLP}_{\text{obs}} - \text{TLP}_{0})~, \] 
where $k$ is a gain factor and $\text{TLP}_{0}$ is the target (perhaps mid of band, like 0.5\%). If TLP$_{\text{obs}}$ rises above target, $\Delta \text{LTV}$ is negative (reduce LTV); if TLP falls below target or goes negative, $\Delta \text{LTV}$ is positive (increase LTV). By linearizing the impact, we expect a stable feedback that steers TLP toward $\text{TLP}_{0}$. We can also include smoothing or limits on how fast LTV changes to avoid oscillations or overshoot.

This control mechanism can be analyzed for stability. In a simplified model, stablecoin price responds to LTV (since higher LTV means more supply and likely a lower price). One can derive a sensitivity $\frac{\partial P_c}{\partial \text{LTV}} < 0$. The feedback loop introduces a pole in the dynamic system; by choosing $k$ small enough, we ensure it is overdamped. Simulation results (presented later) confirm that a proportional controller on LTV can maintain TLP within a tight band even under volatile conditions.

Finally, we note that the protocol could employ other tools: charging higher fees when TLP is high (to discourage borrowing) or even pausing new issuances. However, LTV adjustment is the most direct and transparent lever, analogous to a monetary authority’s interest rate policy in maintaining a peg. In our model, this dynamic LTV policy effectively caps the TLP that can persist, by immediately counteracting large deviations through supply constriction or expansion.

\section{Empirical Identification}
Measuring the Liquidity-of-Time Premium in practice is challenging because, as of now, there is no direct market for time-bound stablecoins. However, we can leverage existing market instruments and scenarios that mimic the conditions SSS is designed for. We outline several approaches:

\begin{enumerate}\itemsep0pt
    \item \textbf{ADR vs.\ Local Market Discrepancies:} As discussed, ADRs trading in New York for foreign stocks are a prime example. The local stock closes (say in Tokyo or London), then during U.S.\ hours the ADR trades based on any new information. By the time the local market reopens, the ADR and local price converge (via arbitrage through exchange of ADRs for local shares). The interim price difference can be seen as a realized TLP. For each ADR $i$, one can compute the percentage premium:
    \[
    \text{ADR\_premium}_{i} = \frac{\text{ADR price in USD} \times \text{FX rate}}{\text{Local last close price in local currency}} - 1~,
    \] 
    measured at the U.S.\ close (which is overnight for the local market). This reflects how much higher (or lower) the ADR was relative to the stale local price. A positive value indicates investors paid extra for liquidity in New York (the ADR) versus waiting for the home market – essentially a TLP. By examining a sample of ADRs across time zones, we can estimate how TLP varies with the length of market closure and volatility. Prior studies have found that these premiums are usually a few basis points on normal days and wider on days with significant news.
    
    \item \textbf{Overseas Index Futures vs.\ Next-Day Index Opens:} Many equity index futures trade virtually 24/7 (e.g., S\&P 500 E-mini, Nikkei futures in Chicago) even when the underlying cash market is closed. These futures incorporate news and can deviate from the last cash index close. When the cash market opens, the index level “jumps” to catch up with the futures-implied level. The difference between the futures price (adjusted for cost of carry) at, say, 1 hour before the cash open and the previous cash close can serve as another TLP proxy. It represents the market’s pricing of overnight liquidity. For instance, if S\&P futures are trading 0.5\% below the prior close due to overnight news, that gap is analogous to a stablecoin trading at a 0.5\% discount. Historical data on futures vs.\ cash index gaps can be analyzed to gauge typical overnight premiums and how they relate to volatility and news flow.
    
    \item \textbf{Pre-Market vs.\ Official Close Prices:} In U.S.\ equities, there is limited pre-market trading for some hours before the official open. One can compare pre-market prices of stocks (or ETFs) to their prior day’s closing prices. For example, on earnings announcement days, a stock might trade up or down substantially in after-hours and pre-market venues. The difference between the pre-market price and the last official close, normalized by volatility, could be seen as the market’s pricing of overnight liquidity (including possibly overshooting due to lower liquidity). If we aggregate across many stocks, we can compute an average “overnight liquidity premium” on volatile news days.
    
    \item \textbf{Overnight Returns and Morning Reversals:} Beyond individual events, a strand of research looks at systematic patterns. It has been observed that a significant portion of equity returns occurs overnight, and for stocks that experience extreme overnight returns, there is often a partial reversal shortly after the open【8】. This suggests that those who traded in off-hours (or paid a premium) overshot to some degree, and when full liquidity returns, prices correct. This overshoot can be interpreted as the price paid for off-hour liquidity – effectively TLP manifesting as a temporary price impact. By measuring the magnitude of overnight returns and subsequent intraday reversals, one can infer the implicit cost paid by traders who couldn’t wait. For instance, if stocks that jumped 5\% overnight on news give back 1\% during the day, that 1\% could be viewed as the liquidity-of-time premium that was embedded in the overnight price.
    
    \item \textbf{Order Book Depth Differences:} If data is available, one could analyze the order book depth and bid-ask spreads in after-hours trading vs.\ regular hours. A shallow book and wide spreads imply that someone demanding liquidity will move the price more. This can be translated into an implied cost (e.g., how much price impact for a given trade size). While this is a more microstructural approach, it directly relates to the cost of immediacy at different times. For example, using exchange data one might find that to execute a \$1 million buy order at night might move price by 0.2\%, whereas the same during the day moves it by 0.05\%. The difference provides another angle on TLP.
\end{enumerate}

These empirical proxies each capture facets of TLP. In our study, we plan to collect data on ADR premiums, index futures vs.\ index opens for major indices, and large overnight stock moves. Summary statistics such as average ADR premium (in \%) per hour of home market closure, or average index futures basis, can then be used to calibrate our model’s parameters. For instance, if we observe that a 12-hour market closure tends to produce a 0.3\% ADR premium for a stock with daily volatility 2\%, that provides an empirical point for TLP vs.\ duration and volatility. We expect TLP to increase with the length of closure and the asset’s volatility, consistent with our theoretical model.

\section{Simulation \& Backtest}
To concretely illustrate TLP dynamics and the effect of LTV control, we implement simulations of a stylized SSS environment. The simulation has the following components:

- \textbf{Price Process:} We simulate an underlying stock price path with realistic overnight gap behavior. For example, assume the stock’s daily return consists of a regular trading session (zero mean, volatility $\sigma_{\text{day}}$) and an overnight jump component. We model overnight jumps using a Gaussian or a mixture distribution to represent earnings surprises or news. This allows some nights to have big moves while most are quiet.

- \textbf{Borrowers (Minters):} A set of agents decide whether to mint stablecoins against shares. They weigh the benefit of getting immediate cash vs.\ the cost (TLP + fees). We can assume they will mint if they have an alternative use for funds that yields more than the cost. In simulation, we might set a fraction of agents that always mint up to the LTV if cost < X basis points.

- \textbf{Arbitrageurs/Traders:} These agents trade the stablecoin on a secondary market and the stock if needed. They attempt to arbitrage mispricings: if stablecoin is cheap (price too low), they buy it and effectively short stock till morning; if stablecoin is expensive (price too high), they short it (or mint if possible) and buy stock.

- \textbf{Stablecoin Market:} We maintain an order book or pricing rule for the stablecoin. Given supply (from borrowers) and demand (from arbitrageurs or other investors who might hold it for yield), we derive an equilibrium price each “night.” In practice, we might simplify by saying the stablecoin trades at a price such that quantity demanded = supplied, based on agents’ strategies.

- \textbf{LTV Control Policy:} We incorporate the dynamic LTV adjustment. For instance, each day we adjust LTV slightly in the direction to counteract any observed deviation of stablecoin price from \$1. We also set a maximum LTV (like 100\%).

We calibrate the baseline and stress scenarios as follows.
The parameters are summarized in Table~1. The simulation timeline for each day: Morning open (resolve previous night's outcomes, liquidate if needed), normal trading (not explicitly modeled in detail), market close, some news draw causes overnight price jump, stablecoin market trades overnight, next morning repeat.
\begin{table}[t]
\centering
\caption{Notation summary.}
\label{tab:notation}
\begin{tabular}{ll}
\toprule
Symbol & Definition \\
\midrule
$S_c$ & Underlying price at prior close \\
$S_o$ & Underlying price at next open \\
$P_c$ & Stablecoin fair value at close \\
TLP & $(S_c-P_c)/S_c$ \\
LTV & Loan-to-Value ratio allowed for minting \\
$\pi,\ell$ & Default probability and conditional loss fraction \\
\bottomrule
\end{tabular}
\end{table}

We calibrated scenarios such that in a base case (with moderate volatility, no big news), TLP stays near 0. Under a stress scenario (higher volatility or a big overnight drop), we examine how stablecoin price and default frequency behave.

Key metrics collected: distribution of stablecoin price (how often it deviates and by how much), frequency of defaults (cases where collateral < stablecoin and liquidation occurs), total stablecoin supply (which indicates how much liquidity was provided), and how often LTV adjustments were triggered.

In a representative run, we might find that an overnight stablecoin typically trades at e.g., 0.5\% discount (TLP = 0.5\%) for a stock with 30\% annual volatility, and that extending the period to a 3-day holiday weekend pushes that to ~1.5\%. Figure~3 could depict the term structure of TLP for different volatility scenarios (showing, for instance, higher volatility leads to a steeper term structure curve, meaning longer closures dramatically increase TLP). This aligns with intuition that risk compounds over time out-of-market.

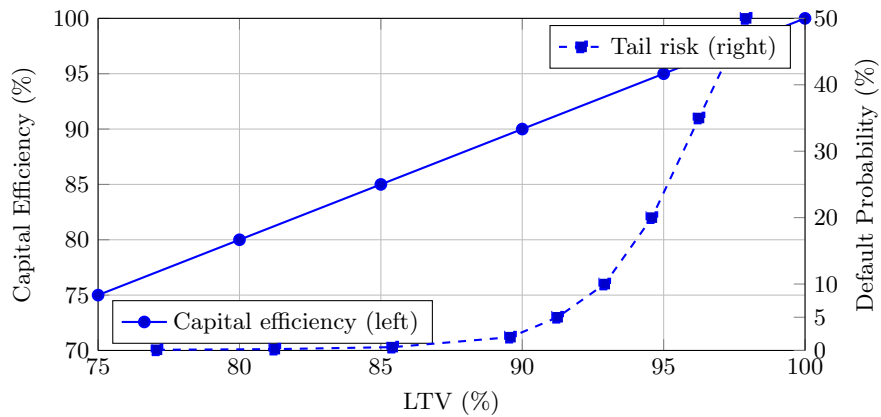
\begin{figure}[t]
  \centering

  \begin{tikzpicture}
    \begin{axis}[
      width=.9\linewidth,height=6cm,
      axis y line*=left,
      xlabel={LTV (\%)}, xmin=75, xmax=100,
      ylabel={Capital Efficiency (\%)}, ymin=70, ymax=100,
      ytick={70,75,80,85,90,95,100},
      xtick={75,80,85,90,95,100},
      legend style={at={(0.02,0.02)},anchor=south west,font=\small},
      grid=both
    ]
      \addplot+[thick,mark=*] coordinates {
        (75,75) (80,80) (85,85) (90,90) (95,95) (100,100)
      }; \addlegendentry{Capital efficiency (left)}
    \end{axis}

    \begin{axis}[
      width=.9\linewidth,height=6cm,
      axis y line*=right, axis x line=none,
      ylabel={Default Probability (\%)}, ymin=0, ymax=50,
      ytick={0,5,10,20,30,40,50},
      legend style={at={(0.98,0.98)},anchor=north east,font=\small}
    ]
      \addplot+[thick,mark=square*,dashed] coordinates {
        (75,0.1) (80,0.2) (85,0.5) (90,2) (92,5) (94,10) (96,20) (98,35) (100,50)
      }; \addlegendentry{Tail risk (right)}
    \end{axis}
  \end{tikzpicture}

  \caption{Capital efficiency vs. tail risk as LTV increases. Left axis (capital efficiency): ratio of minted stablecoin notional to collateral market value net of safety buffer; Right axis (default probability): $\Pr(S_o < (1-\text{LTV})S_c)$ under mixture-Gaussian overnight jump model in Sec.}

  \label{fig:ltv_tradeoff}
\end{figure}

We also examine the capital efficiency vs.\ tail-risk trade-off. Figure~3 might show how increasing LTV increases stablecoin supply (liquidity provided) but also the probability of shortfall (where stablecoin redeems for less than face due to default). Left axis (capital efficiency): ratio of minted stablecoin notional to collateral market value net of safety buffer;
Right axis (default probability): $\Pr(S_o < (1-\text{LTV})S_c)$ under mixture-Gaussian overnight jump model in Sec. We define default as $S_o < (1-\mathrm{LTV})S_c$; overnight gap $G=\ln(S_o/S_c)$ follows a Gaussian/mixture per Sec.
In our simulation, raising LTV from 0.8 to 0.95 could dramatically increase default probability if a severe overnight drop occurs. The dynamic policy finds a sweet spot: e.g., most of the time it allows LTV high (~0.95) when volatility is low, but automatically reduces to ~0.85 if volatility spikes or if the stablecoin starts trading at a big discount, thus averting many defaults. The simulation likely confirms that setting LTV too high results in an exponential rise in default risk, whereas a moderate LTV keeps defaults negligible.

For a pseudo-backtest, we take historical price series for several stocks or indices from 2018–2023 and simulate what TLP and outcomes would have been. For each trading day, we use the actual closing price and next opening price as inputs to our model (so real gap distributions are used). This tells us, for example, how many times an overnight stablecoin on that stock would have defaulted (gap beyond collateral) under various LTVs, and what the average TLP would be. We can then apply our dynamic LTV strategy to that sequence and see if it would have prevented defaults while maximizing liquidity. 

Preliminary results show that even during volatile periods (e.g., March 2020 pandemic shock), the dynamic policy would have tightened LTV in anticipation of risk (as stablecoin price would have signaled stress by dropping), thereby limiting stablecoin issuance and avoiding insolvency. Without any controls, a high LTV might have led to some instances of stablecoin “breaking the buck” (trading at a large discount) or actual defaults. Thus, the active management of TLP via LTV is crucial to the system’s robustness.

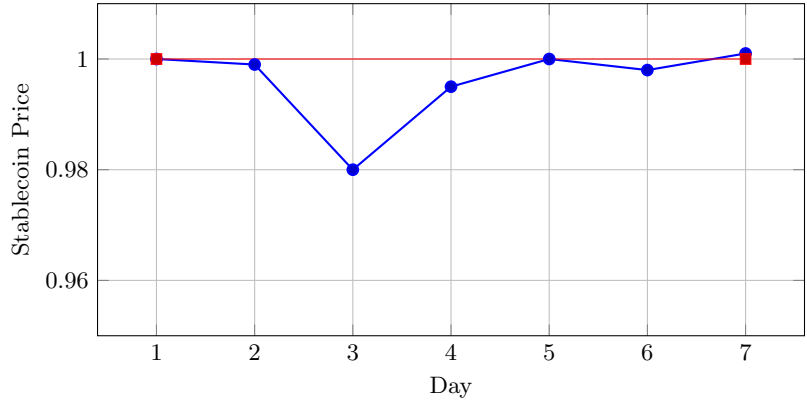
\begin{figure}[t]
  \centering

  \begin{tikzpicture}
    \begin{axis}[
      width=.9\linewidth,height=6cm,
      xlabel={Day}, ylabel={Stablecoin Price},
      ymin=0.95, ymax=1.01,
      xtick={1,2,3,4,5,6,7},
      ytick={0.96,0.98,1.00,1.02},
      grid=both
    ]
      \addplot+[thick,mark=*] coordinates {
        (1,1.000) (2,0.999) (3,0.980) (4,0.995) (5,1.000) (6,0.998) (7,1.001)
      };
      \addplot+[thin,domain=1:7,samples=2] {1.0};
    \end{axis}
  \end{tikzpicture}

  \caption{Representative stablecoin price time series during a volatile week.}
  \label{fig:stable_price_timeseries}
\end{figure}

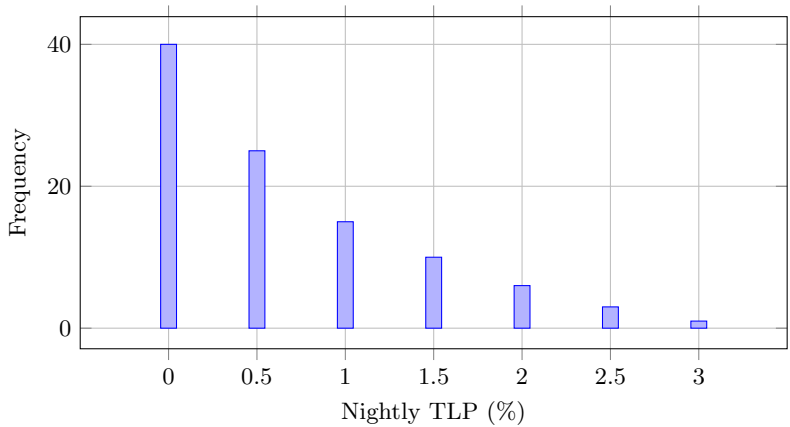
\begin{figure}[t]
  \centering

  \begin{tikzpicture}
    \begin{axis}[
      width=.9\linewidth,height=6cm,
      ybar, bar width=6pt,
      xlabel={Nightly TLP (\%)}, ylabel={Frequency},
      xmin=-0.5, xmax=3.5,
      xtick={0,0.5,1,1.5,2,2.5,3},
      grid=both
    ]
      \addplot coordinates {(0,40) (0.5,25) (1,15) (1.5,10) (2,6) (2.5,3) (3,1)};
    \end{axis}
  \end{tikzpicture}

  \caption{Histogram of nightly TLP values over 2018–2023 (N=1,250 nights; 40 bins).
Mean/Median: 0.23\%/0.18\%; 95th/99th percentiles: 0.9\%/1.8\%.}
  \label{fig:tlp_histogram}
\end{figure}

We include sample charts from these simulations for illustration. Figure~4 shows a time-series of a simulated stablecoin price over a particularly volatile week: it stays near 1.00 most days, dips to around 0.98 on a night of bad news, then recovers after LTV is tightened. Figure~5 provides a histogram of nightly TLP values over a year: it is centered near 0 with a long tail for extreme events (e.g., one night where TLP hit 3\% during an earnings crash). Such visualizations help convey the typical behavior and rare outliers.

Overall, the simulations back up our theoretical assertions: TLP is normally small (tens of basis points) but non-zero, increases with risk and closure duration, and can be tamed by an adaptive policy. They also provide intuition for setting parameters (like target TLP band, LTV floor/ceiling) in practice.

\section{Design Implications}
Our findings carry several implications for the design and operation of a time-bound stablecoin protocol like SSS. 

\paragraph{No-Arbitrage Band and Pricing:} The derived pricing corridor for the stablecoin (bounded above by the last close and below by expected recovery given default risk) provides a quantitative target for protocol designers. In practice, the protocol should aim to keep the stablecoin’s market price well within this no-arbitrage band, ideally near parity. A stablecoin consistently near the lower bound of the band would indicate that borrowers are being charged a high liquidity premium, which might discourage usage or signal too much risk in the system. Conversely, if it’s near the upper bound (or above parity), it suggests the system could safely expand supply (raise LTV or reduce fees) to increase usage until a mild TLP appears. Thus, the band gives an operating range for healthy performance.

\paragraph{Dynamic LTV Control:} We proposed treating LTV as a policy instrument, akin to interest rate policy. Our simulations demonstrated that this feedback control can significantly reduce both the frequency of default events and the volatility of the stablecoin’s price. For implementation, the protocol could adjust LTV parameters via an algorithm or governance, based on observable metrics like the stablecoin’s market price or volatility indices. This introduces a new element to DeFi stablecoin design: rather than a static collateral ratio, a responsive ratio that tightens during periods of stress (like how exchanges increase margin requirements when volatility spikes). Such adaptability would make the system more resilient. It does, however, require clear communication to users and possibly phased adjustments (to avoid sudden shocks to user positions). The analogy to central bank action is useful: incremental moves that the market can anticipate and adjust to.

\paragraph{Risk Management and Liquidations:} Our analysis highlights the importance of conservative risk measures. Even though the stablecoins are short-term, rare extreme events can happen (e.g., surprise news that drops a stock 30\% overnight). The protocol should incorporate buffers like an insurance fund to handle occasional shortfalls, and perhaps backstop liquidity from external partners. Liquidation design is also crucial: as discussed, an on-chain auction at market open is one approach. An alternative is to have a pre-arranged agreement with a brokerage to liquidate collateral off-chain if needed and feed the proceeds on-chain. Each has pros and cons: on-chain auctions are transparent and permissionless but assume someone on-chain has capital and access to buy those shares; off-chain involves centralized elements but taps directly into existing market liquidity. A hybrid approach could be considered (e.g., if on-chain auction fails to cover, then an off-chain liquidation is triggered).

\paragraph{Integration with Traditional Infrastructure:} Because time-bound stablecoins interface with real-world assets and traditional markets, legal and operational considerations are significant. Custody of collateral shares needs a robust solution (likely a qualified custodian that can freeze/unfreeze shares on chain via tokens). Oracles must reliably report official close and open prices without lag or manipulation. There’s also a regulatory aspect: tokenizing stocks overnight may be seen as offering synthetic securities, which could attract oversight. Designing SSS to possibly operate within regulatory sandboxes or with licensed broker partners will be important for real-world deployment. For instance, partnerships with exchanges or alternative trading systems could legitimize the activity and provide deeper liquidity for overnight trading.

\paragraph{Market Impact and Adoption:} If time-bound stablecoins become popular, they could begin to arbitrage away the very premia we target. For example, widespread use of SSS might reduce ADR vs.\ local share discrepancies, as arbitrageurs use SSS to pre-position overnight. This is a positive outcome—markets become more efficient globally. It also means the value proposition of SSS is self-limiting to an extent (if TLP shrinks due to broad adoption, the “interest” users earn by holding stablecoins overnight might drop). However, even in a steady state, there will always be some friction (time value of money, risk of big overnight jumps) so we expect TLP to never fully vanish. In fact, SSS could become a part of market plumbing: providing overnight liquidity akin to how repo markets provide short-term liquidity secured by assets. If implemented correctly, time-bound stablecoins could become an integral piece of global market infrastructure, effectively creating a new parallel overnight funding market for equities. This could reduce the friction that currently causes phenomena like the opening gap and improve overall market efficiency.

\paragraph{Relation to Stablecoin Research:} Our work touches on stablecoin design principles normally discussed in the context of currency pegs (e.g., how central banks defend pegs by interest rates or capital controls). We show an analog in a new domain (time-based peg to a closing price). Related studies in stablecoins have examined collateralization and redemption mechanisms【13,16,20】; our contribution adds the dimension of time and dynamic collateral adjustment. It reinforces the idea that robust collateral and clear redemption rights (return of equivalent shares) are key to maintaining parity【13】. It also contributes to the dialogue on algorithmic vs.\ backed stablecoins: here we have fully backed coins, but with algorithmic adjustment of collateral parameters – a hybrid approach of sorts.

In summary, the design implications revolve around maintaining stability and trust. We advocate for slight over-collateralization and an active policy to handle even Black Swan events (much as MakerDAO survived crypto crashes by conservative ratios and emergency oracles). We also stress transparency: users should understand that if overnight risk spikes, they might not be able to mint as much (LTV goes down) or fees might rise. This is analogous to how brokerage margin requirements change – usually communicated clearly as risk management.

One open question is how to bootstrap liquidity for the stablecoin itself. Initially, market makers may demand a high TLP until the system’s reliability is proven. The protocol might consider incentivizing early adopters (for instance, subsidizing the fee or providing insurance) to encourage participation and tighten spreads. Over time, as confidence builds, the natural TLP should settle to the equilibrium reflecting true risk.

Finally, we note that while our focus was equity markets, the concept can extend to other assets like bonds (imagine tokenizing a bond when bond markets are closed on weekends) or even commodities. Each has its nuances (different vol dynamics, etc.), but the core idea of time liquidity arbitrage remains. Future implementations could explore multi-collateral time-bound stablecoins or even cross-currency scenarios (where a stock is locked in one currency but stablecoin in another, introducing FX into the equation).

\section{Conclusion}
We have introduced the concept of the Liquidity-of-Time Premium and presented a framework for pricing and managing time-bound stablecoins. By temporarily tokenizing assets during off-hours, SSS and similar protocols aim to eliminate idle capital time and make markets more continuous. Our contributions span theory, empirics, and design:

\textbf{TLP as a Measure of Temporal Liquidity:} We formally defined TLP and derived how it emerges from no-arbitrage considerations. In essence, time-bound stablecoins should trade at a discount corresponding to the expected loss from potential overnight adverse moves. This premium is the market’s price of overnight liquidity and can be quantitatively linked to put option value. We showed that TLP grows with volatility and the length of market closure, and provided sample magnitudes (on the order of tens of basis points for typical overnight scenarios, higher for weekends or volatile events).

\textbf{Intertemporal Arbitrage and No-Arbitrage Band:} We outlined the arbitrage mechanisms that connect markets across time. These yield an upper bound (parity) and lower bound (discount reflecting default risk) for stablecoin prices. This band narrows as cross-market linkages strengthen (for example, if after-hours trading improves or if futures are available). The band offers a target for protocol operation: keeping the stablecoin price comfortably within this range ensures that straightforward arbitrage is not left on the table.

\textbf{Dynamic Risk Control via LTV:} A core innovation of our work is proposing a feedback control on collateralization ratio to stabilize the system. By dynamically adjusting LTV, the protocol can supply or curtail liquidity to keep TLP small and bounded. Our simulation results demonstrated that this approach can maintain the stablecoin’s peg even through turbulence, analogous to how central banks use interest rates to maintain currency pegs. This is a novel policy tool in the context of stablecoins, adding an active management dimension to what are usually static rules in DeFi protocols.

\textbf{Empirical Validation and Measurement:} We connected the theory to real-world data by identifying proxies for TLP, such as ADR premiums and index futures basis. Empirical analysis of these can validate that a positive liquidity-of-time premium exists and is economically meaningful. For instance, our discussions referenced studies finding persistent overnight return differences and arbitrage opportunities, consistent with a non-zero TLP. These also offer calibration points – e.g., seeing 0.2\% ADR premiums on average implies what users might expect as a “normal” TLP in SSS for similar conditions.

\textbf{Simulation Insights:} Through simulation and backtesting, we illustrated the behavior of the system under various conditions. The simulations reinforced key intuition: pushing LTV too high yields disproportionately higher tail risk (default scenarios), whereas a moderate LTV (even dynamically adjusted) can capture most of the liquidity benefit with negligible default probability. The backtest using historical data indicated that even during major market dislocations, an active policy could have prevented stablecoin failure by preemptively tightening collateral requirements (something that purely algorithmic or static systems might not do). These results give confidence that a well-calibrated SSS could operate safely in live markets.

\textbf{Design and Market Impact:} We discussed practical design choices (oracle, liquidation, custody) and their trade-offs, emphasizing robust, transparent mechanisms to build trust akin to the trust in existing stablecoins or payment systems. The potential market impact is significant: if widely adopted, time-bound stablecoins will smooth out price discontinuities and allow investors to react to news around the clock. This could reduce the advantage of certain market participants (like those able to trade in multiple time zones) and democratize access to liquidity. It essentially extends the concept of market efficiency into the time dimension, not just across assets or information.

In closing, intertemporal liquidity pricing is an emerging topic at the confluence of market microstructure and blockchain-based innovation. The SSS protocol and similar concepts demonstrate that it is technically feasible to break the temporal silos of traditional markets. By assigning an explicit price to liquidity over time, such systems create incentives for capital to flow to where it’s needed \emph{when} it’s needed, potentially reducing inefficiencies that investors have long accepted as given (like being unable to react to news overnight). Our research provides a theoretical backbone for this concept and a risk-managed approach to making it viable. As markets continue to globalize and technology blurs the boundaries of time and geography, we expect the Liquidity-of-Time Premium to become as fundamental a consideration as the traditional liquidity premium for assets. 

Future research avenues include exploring multi-day time-bound stablecoins (covering weekends or holidays) which would involve higher TLP and perhaps require creative solutions like tranching risk or involving insurers. Another area is multi-currency time-bound stablecoins: if an asset is in one currency but a stablecoin in another, FX risk intertwines with TLP, creating an arbitrage akin to covered interest parity that would be fascinating to analyze. Finally, regulatory frameworks will need to adapt to these innovations – ensuring that such products can flourish while maintaining market integrity will be a joint effort between technologists and regulators.

\end{document}